\documentclass[10pt,journal,compsoc]{IEEEtran}
\usepackage{mathptmx} 
\usepackage{fancyhdr}
\usepackage[normalem]{ulem}
\usepackage[hyphens]{url}
\usepackage[sort,nocompress]{cite}
\usepackage[final]{microtype}
\usepackage[keeplastbox]{flushend}
\usepackage{graphicx}
\graphicspath{ {./images/} }
\usepackage{subfig}
\usepackage{caption}
\usepackage{tabularx}
\usepackage{url}
\usepackage{listings}
\usepackage{color}
\usepackage{lstlinebgrd}
\usepackage{pgffor}
\usepackage[bookmarks=true,breaklinks=true,letterpaper=true,colorlinks,linkcolor=black,citecolor=blue,urlcolor=black]{hyperref}

\definecolor{codegreen}{rgb}{0,0.6,0}
\definecolor{codegray}{rgb}{0.5,0.5,0.5}
\definecolor{codepurple}{rgb}{0.58,0,0.82}
\definecolor{backcolour}{rgb}{0.95,0.95,0.92}
\definecolor{highlight}{rgb}{1.0,0.9,0}
 
\lstdefinestyle{mystyle}{
    backgroundcolor=\color{backcolour},   
    commentstyle=\color{codegreen},
    keywordstyle=\color{magenta},
    numberstyle=\tiny\color{codegray},
    stringstyle=\color{codepurple},
    basicstyle=\small,
    breakatwhitespace=false,         
    breaklines=true,                 
    captionpos=b,                    
    keepspaces=true,                 
    numbers=left,                    
    numbersep=5pt,                  
    showspaces=false,                
   showstringspaces=false,
    showtabs=false,                  
    tabsize=2
}

\lstdefinelanguage{myasm} {
	morekeywords={
		rax,rbx,rcx,rdx,rdi,rsi,rbp,rsp,r8,r9,r10,r11,r12,r13,r14,r15,
	},
	sensitive=true,
	morecomment=[l]{\#},
	morestring=[b]"
}

\newcommand{\bgndcolor}[1]{%
\def\retval{\color{backcolour}}%
\foreach\x in{2,3,6,7,8,9,10,15,16,21,22,23,24,25,26,27,36,37,38,39,40,41,42,43,44,45}%
{\ifnum\x=#1\gdef\retval{\color{highlight}}\breakforeach\fi}%
\retval%
}%

\title{Helper Without Threads: Customized Prefetching for Delinquent Irregular Loads} 
\author{Karthik Sankaranarayanan, Chit-Kwan Lin and Gautham Chinya,\\
E-mail: karthik.sankaranarayanan@intel.com\\
Intel Corporation, 2111 NE 25th Ave, Hillsboro, OR 97124}

\begin{document}
\maketitle
\pagestyle{plain}


\begin{abstract}

The growing memory footprints of cloud and big data applications mean
that data center CPUs can spend significant time waiting for memory.
An attractive approach to improving performance in such centralized
compute settings is to employ prefetchers that are customized per
application, where gains can be easily scaled across thousands of
machines.  Helper thread prefetching is such a technique but has yet
to achieve wide adoption since it requires spare thread contexts or
special hardware/firmware support.  In this paper, we propose an
inline software prefetching technique that overcomes these
restrictions by inserting the helper code into the main thread itself.
Our approach is complementary to and does not interfere with existing
hardware prefetchers since we target only delinquent irregular load
instructions (those with no constant or striding address patterns).
For each chosen load instruction, we generate and insert a customized
software prefetcher extracted from and mimicking the application's
dataflow, all without access to the application source code.  For a
set of irregular workloads that are memory-bound, we demonstrate up to
2X single-thread performance improvement on recent high-end hardware
(Intel Skylake) and up to 83\% speedup over a helper thread implementation
on the same hardware, due to the absence of thread spawning overhead.
\end{abstract}

\section{Introduction}
\label{sec:intro}


The rise of the cloud and big data has caused the memory footprint of
applications to grow faster than the pace of technology scaling (i.e.,
memory capacity and core counts). Moreover, as data parallel workloads
increasingly move away from the CPU into GPUs, FPGAs and accelerators, 
the CPU is faced with a rise of irregular memory applications. As a result, 
data center CPUs can spend a significant fraction of execution cycles waiting 
for the caches~\cite{warehouse_isca15}. Yet, despite the large core counts
available, Amdahl's Law means that mitigating such single-thread
performance bottlenecks remains crucial to achieving improved overall
performance~\cite{brawnycores,treadmill}. Interestingly, the 
centralization of compute in the data center can be seen as an
opportunity to be exploited.  By customizing performance optimizations
\emph{per application}, gains can be scaled across many thousands of
machines.  This approach relies on obtaining intimate knowledge of an
application's behavior through profiling and hardware performance
counters~\cite{topdown}, and using such information to extract optimal
performance from the hardware for the target application.

Speculative precomputation~\cite{helper_dubois98, helper_isca99,
  helper_hpca01, slicing_isca01, helper_isca01, helper2_isca01,
  helper_micro01, tullsen_micro01, helper_micro02,helper_cf04},
otherwise known as helper threading~\cite{helper_asplos02,
  helper_pldi02, helper_asplos04, helper_cgo04, helper_micro05,
  helper_ipdps06, helper_asplos11} is such a technique. It reduces the
single-thread latency of an application by using idle thread contexts
in the hardware to spawn special-purpose, speculative threads called
\emph{helper threads}. Helper threads contain computation extracted
from the main thread and \emph{consume} the latency of execution on
behalf of it.  They encounter cache misses and branch mispredictions
ahead of the main thread, and act as execution-driven prefetchers or
branch predictors for the main application, thereby improving its
latency significantly. Their benefit accrues from the fact that they
are tailored to the specific application they are extracted from, and
therefore orchestrate the hardware precisely to suit its needs.

However, helper threads can be be tricky to implement efficiently.
Although the original ideas appeared over two decades ago, we are not
aware of any commercial processors with hardware support for helper
threads today. Note that industry-strength compiler support is available
for code generation of helper threads on multicore CPUs ~\cite{icc,ibmcc}.
However, the corresponding hardware support for generating low-overhead 
micro/nano threads (for e.g., as in ~\cite{helper_isca99}) is absent. 
Hence, the dynamic thread spawn overhead is still significant in current 
operating systems. In the absence of such specialized hardware support, 
helper threads have two disadvantages today: (1) the need for spare 
thread contexts; and (2) the difficulty in synchronizing and match the rates of the main
application and the helper threads.  In this work, we overcome both
these limitations with an inline prefetching technique inspired by
software pipelining~\cite{rau_micro81,lam_pldi88}, yet our method
retains an important benefit of helper threads: it works without
access or modification to the application source code.  This makes our
technique attractive to cloud service providers who run third-party
applications at scale.

\subsection{Delinquent Irregular Loads}


Load instructions in a program can fall into three categories: (a)
constant address, (b) striding, and (c) irregular.  Constant address
loads are loads whose virtual address does not change over multiple
dynamic instances of the load (for e.g., global variables and stack
accesses).  Striding loads are those with successive virtual addresses
following an arithmetic progression (for e.g., array accesses).
\emph{Irregular loads} are those which do not fall into either of the
above two categories (for e.g., indirect and pointer references).
Furthermore, loads that frequently miss in the cache are said to be
\emph{delinquent}.  While current hardware mechanisms are effective at
prefetching regular address patterns~\cite{prefetch_primer},
\emph{delinquent irregular loads} (DILs) remain a challenge.

\begin{figure}[h]
    \centering
    \includegraphics[width=\columnwidth]{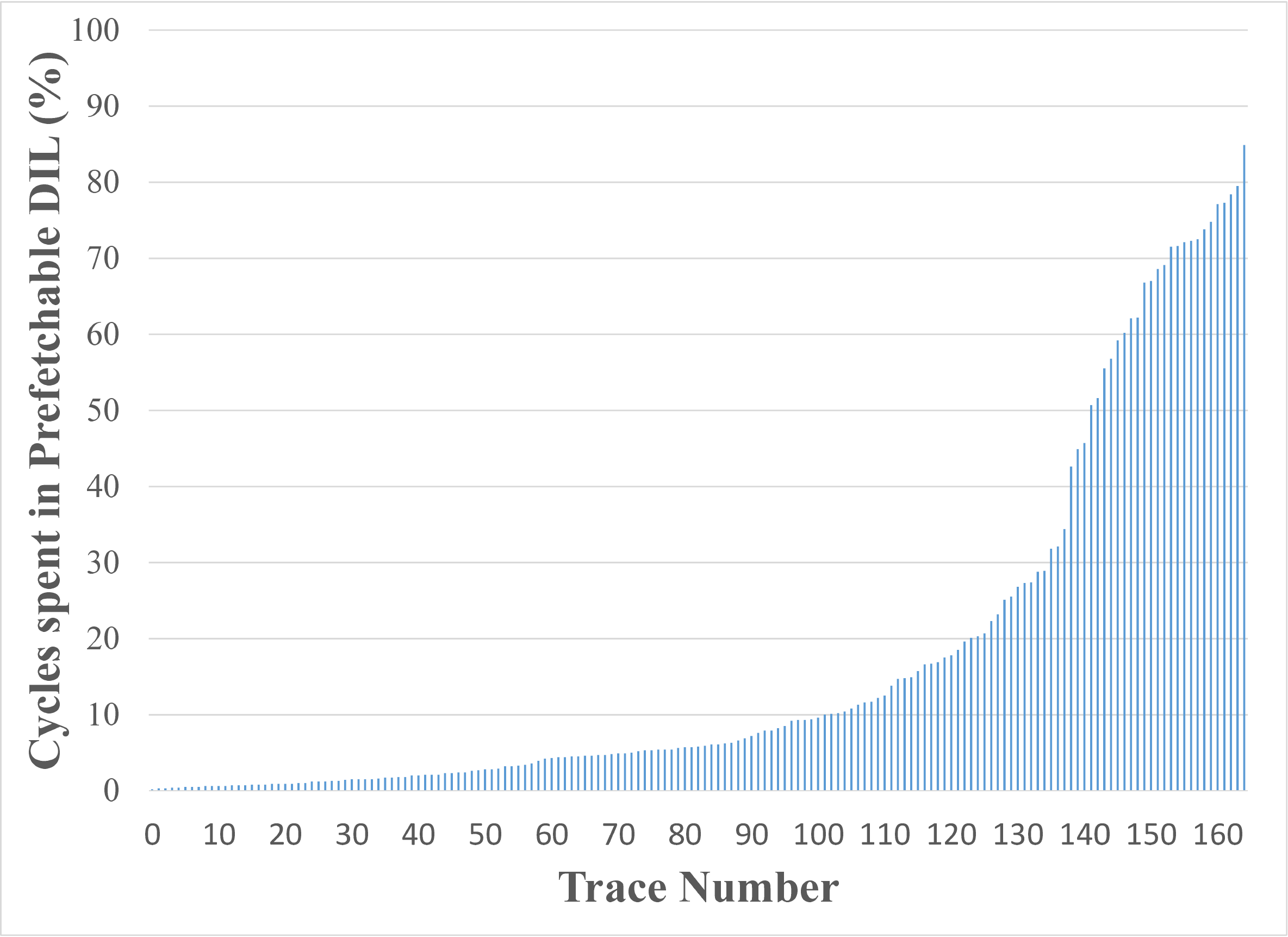}
    \caption{Fraction of CPU cycles spent waiting for data by specific
      memory-bound, prefetchable delinquent irregular loads (DILs).}
    \label{fig:tracedata}
\end{figure}

For a set of 165 traces from commercial applications identified by our
framework as containing prefetchable DILs (see
Section~\ref{sec:method} for a formal definition),
Figure~\ref{fig:tracedata} shows the fraction of CPU cycles spent
stalled waiting for data at the time of retirement of the specific
DILs. The traces are from several client applications (productivity, 
games, content creation), server applications (cloud, database, enterprise, 
HPC) and CPU benchmarks simulated on a cycle-accurate simulator modeling 
the Intel Skylake~\cite{skylake} microarchitecture. On average, each trace 
has about three prefetchable DILs that
are memory-bound.  Should all that stall time be reduced to zero, the
potential geometric mean speedup possible in these traces is 15\%,
which is significant headroom.  However, these opportunities were identified
from a universe of over 2000 traces and building a prefetcher in hardware
for such a narrow focus is not a profitable \emph{microarchitectural} trade-off
since such a prefetcher will be unused most of the time. On the other hand, 
large silicon and software companies currently employ significant \emph{software}
resources for manually optimizing select applications (e.g., 
Figure~\ref{fig:runnable} is from a search engine). The scale of these 
high-value applications justifies the extra effort spent in optimizing 
them (\cite{asmdb_isca19,softsku_isca19}). As long as these applications
fall into the right side of figure~\ref{fig:tracedata}, targeting the DILs 
from them through software is a better strategy than implementing a 
specialized prefetcher in hardware. Hence, we propose an inline software 
prefetching technique that selectively targets DILs that are memory-bound 
for prefetching.

\begin{figure}[h]
    \centering
    \includegraphics[width=\columnwidth]{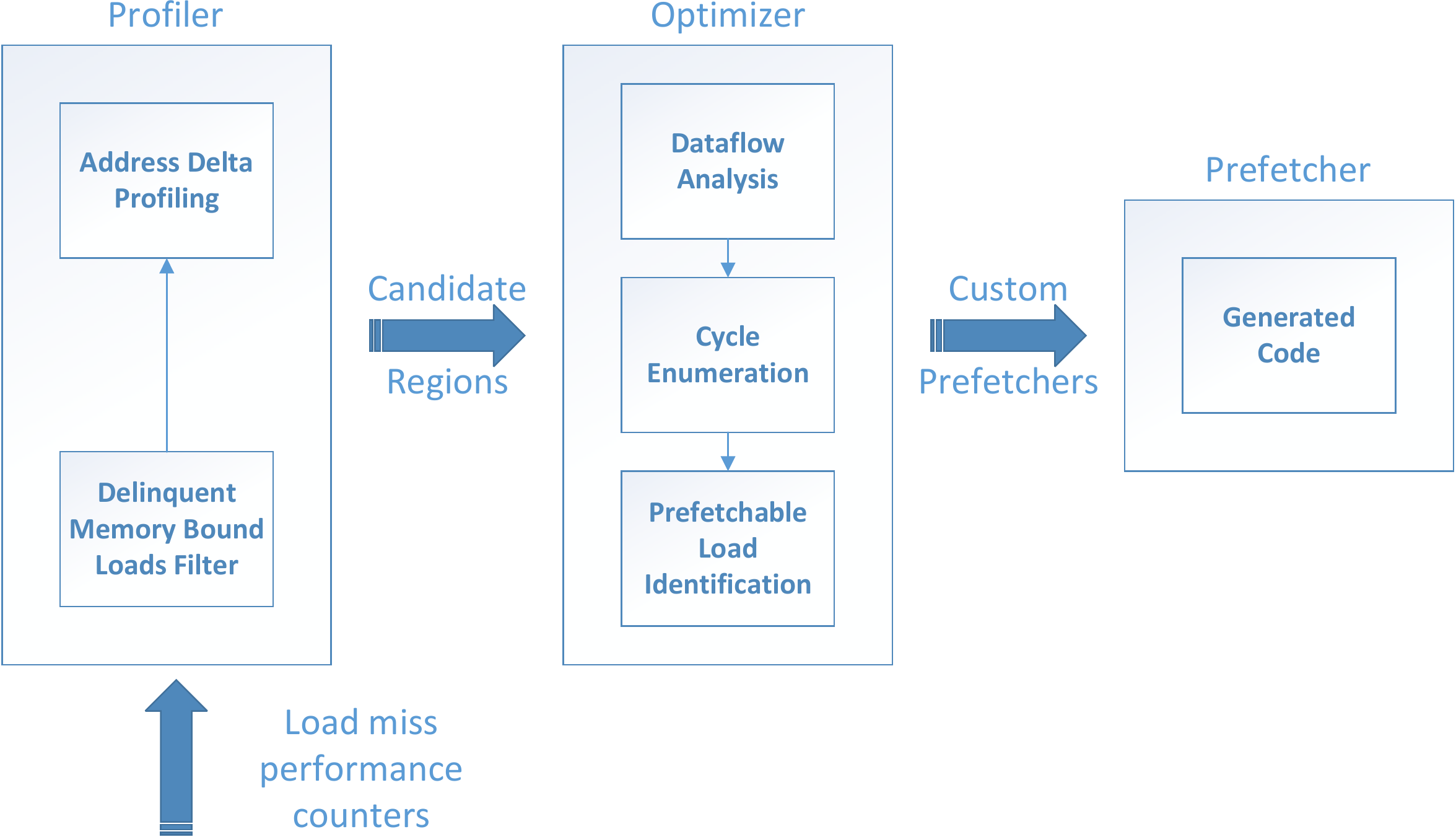}
    \caption{An overview of our customized prefetching approach, which
      requires no access to program source code.  
    }
    \label{fig:blockdiagram}
\end{figure}

Figure~\ref{fig:blockdiagram} summarizes our method.  Prefetchers
(either in hardware or in software) must meet accuracy, timeliness,
and bandwidth criteria; specifically, an issued prefetch must be to
the correct address just ahead of the actual demand, and must not
throttle demand loads by consuming too much memory bandwidth.  When
these criteria are not met, the cache is polluted and performance
suffers.  Software prefetchers face an additional challenge of
unintended interactions with hardware
prefetchers~\cite{software_prefetching}.  We expressly designed our
software prefetcher to avoid these complications by placing an
emphasis on the selection of the load instructions that we prefetch:
we only target loads that are extremely difficult for the hardware to
prefetch, namely, memory-bound DILs.  Through detailed profiling and
dynamic dataflow analysis~\cite{dynamic_slicing}, our method identifies
candidate memory-bound DILs that are part of inner loops and are likely
to benefit from prefetching with minimal software prefetcher
complexity.  We call such DILs {\em prefetchable} and generate
\emph{customized prefetching code} for each.  Similar to helper
threads~\cite{helper_cgo04,helper_asplos04}, the prefetching code is extracted from
the dataflow of the application itself.  However, unlike helper threads
(which require either spare thread contexts or special hardware support 
to run), our method inserts the prefetcher code into the application 
machine code directly.  We find that the customized prefetching 
code sequence is usually small and its overhead of implementation is 
negligible compared to the significant performance gain we
achieve with improved prefetching.

\subsection{Contributions}
This paper makes the following contributions:
\begin{itemize}
\item We overcome the limitations of helper threads with
  an inline prefetching scheme that does not require 
  spare thread contexts or special hardware support; 
\item We eliminate the need for thread synchronization in helper
  prefetching by employing a statically-controlled prefetch distance
  and a prefetcher generation process inspired from software
  pipelining;
\item For a set of irregular memory workloads, we demonstrate up to 2X 
  end-to-end execution time speed-up on current high-end hardware and
  up to 83\% gain over helper threads. 
\end{itemize}


\section {Related Work}
\label{sec:relwork}

A full discussion of the literature on prefetching is beyond the scope
of this paper; the interested reader is referred to Falsafi and
Wenisch~\cite{prefetch_primer} and Lee {\em
  et. al.}~\cite{software_prefetching} for thorough treatments of the
hardware and software approaches to prefetching and the various
challenges involved. Limiting our focus to irregular load prefetching,
prior work can be classified into three main categories:

\textbf{Microarchitectural techniques} use on-chip
storage to record patterns in the addresses of the irregular load and
predict a future address to prefetch if the current address is from
one of the recorded patterns~\cite{imp, avd_micro05, jain_isb, temporal_micro19, ghb,
  markov_prefetcher, dbp, ml_prefetcher}.  These approaches require
large on-chip storage, the cost of which continues to preclude their
commercial viability. 
The Indirect Memory Prefetcher (IMP)~\cite{imp} is an exception---it
uses very little on-chip storage by targeting specific indirect memory
patterns of the form \(a[b[i]]\), where the array \(a\) is addressed
by a striding feeder load \(b[i]\). At runtime, IMP records in a
hardware table the relationship between the striding load and the
irregular load address and uses this to predict future addresses.  The
goals of our present work are to minimize prefetcher implementation
complexity, as well as improve performance on current hardware.
Hence, we choose the software implementation route.  Moreover, as
described in Section~\ref{sec:example}, our dataflow analysis
framework is generic enough to handle more complex patterns such as
\(a[f(b[i])]\) where \(f\) is any arbitrary function.

\textbf{Computation-based techniques} typically execute program
instructions ahead of time to prefetch delinquent loads.  While
computation-based prefetching~\cite{slipstream, helper_cgo04,
  runahead, dependent_misses} can be accurate, large runahead buffers
or spare thread contexts for running helper threads are
resource-intensive, especially considering their energy costs.  Since
regular loads are the overwhelming majority in general purpose
applications, dedicating special hardware resources to handle
comparatively rare events such as irregular loads may not represent a
good microarchitectural trade-off.  In contrast, a software
implementation is more flexible and can be invoked to incur the cost
only when the benefit is known to be greater.

There are other limitations to computation-based methods beyond just
the cost of implementation.  For example, runahead~\cite{runahead} is
a technique that ignores a branch misprediction and continues
execution to extract prefetching benefit from control independent
instructions.  However, it is not designed to handle dependent misses
(i.e., misses whose addresses are data-dependent on previous misses).
Our approach handles dependent misses by prefetching the entire
dependency chain through to the missing leaf instruction (see
Section~\ref{sec:example}).

Helper
threads~\cite{helper_dubois98,helper_isca99,helper_hpca01,slicing_isca01,helper_isca01,helper2_isca01,helper_micro01,tullsen_micro01,helper_micro02,helper_cf04,helper_asplos02,helper_pldi02,helper_asplos04,helper_cgo04,helper_micro05,helper_ipdps06,helper_asplos11}
extract the backward slice of a delinquent load and run it on a spare
thread context.  When the latency of the backward slice is less than
that of the original loop, the helper thread runs ahead of the main
thread and prefetches memory accessed by the main thread into the
cache. This technique has the advantage of being flexible enough to be
implemented in
hardware~\cite{helper_dubois98,helper_isca99,helper_hpca01,slicing_isca01,helper_isca01,helper2_isca01,helper_micro01,tullsen_micro01,helper_micro02,helper_cf04,decoupled_micro08},
or
software~\cite{helper_asplos02,helper_pldi02,helper_asplos04,helper_cgo04,helper_micro05,helper_ipdps06,helper_asplos11}.
It can also work in the absence of high-level source code and has been
demonstrated in a compiler~\cite{helper_asplos02}, binary
tool~\cite{helper_pldi02}, or a dynamic
optimizer~\cite{helper_micro05}. However, all prior work in this area
has either required spare thread contexts or special hardware/firmware
support.  Virtual Multithreading (VMT)~\cite{helper_asplos04}
overcomes the need for spare thread contexts by partitioning the
registers available to the compiler between the main and the helper
computation.  However, it still requires special \emph{yield}
instructions to orchestrate the transfer of control between the
virtual threads and corresponding modifications to the processor
firmware.  In our work, by choosing an inline implementation, we (1)
avoid the need for any extra hardware or firmware support; (2)
sidestep thread spawning overheads and synchronization bugs since
there are no threads to run; and (3) make straightforward the rate
matching between the main computation and the prefetcher by statically
setting the prefetch distance.

Similar to our approach, recent work~\cite{prefetching_asplos20} has 
proposed inserting prefetch hints based upon binary analysis. However,
due to the absence of control flow analysis, it requires specialized 
hardware support in the form of speculative loads. Moreover, the 
benefits demonstrated are on top of a simulated microarchitecture 
without state-of-the-art hardware prefetchers such as \cite{vldp,imp}. 
In contrast, we show benefit on existing hardware (not requiring
special hardware support) and in comparison with state-of-the-art hardware 
prefetchers. 

\textbf{High level language software techniques.} Broadly speaking,
software-based prefetchers are typically concerned with inserting
pre\-fetch hints~\cite{dynamic_injection_pldi04} into a program or
modifying its data structures.
Most~\cite{compiler_prefetch_asplos96,indirect_asplos18,jump_pointer,software_prefetching,chilimbi_pldi02}
rely on access to the program's source code.  For instance, Roth and
Sohi~\cite{jump_pointer} augment the data layout of linked data
structures with a jump pointer that acts as a prefetch pointer.
Others~\cite{ainsworth_cgo17,indirect_asplos18} have demonstrated
significant speedup with programmable prefetching.  However, many
real-world situations preclude access to program source code, e.g.,
while using third party libraries or when serving third-party
applications in the cloud.  In such situations, the ability to improve
performance in the absence of source code is attractive.
This work retains such capability from its lineage in helper
threads. Finally, while our prefetcher generation is inspired from
software pipelining~\cite{rau_micro81,lam_pldi88,modulo_micro97}, it
is not a static instruction scheduling technique. Our target is
performance improvement over dynamically scheduled out-of-order
processors that hold multiple iterations of loops in their instruction
window. The performance improvement is exclusively due to the
duplication of code that stays a constant number of iterations ahead.

\section {A Motivating Example}
\label{sec:example}

Let us now consider an example scenario where memory-bound DILs occur
frequently.  Hash tables are widely used because of their algorithmic
efficiency in converting expected linear and logarithmic time
operations into expected constant time operations~\cite{clr}.  For
instance, they are used to implement associative arrays in popular
scripting languages such as Python and R, and in relational databases
for indexing.  However, the underlying hash functions that generate
hash table keys are designed specifically to disrupt data locality,
{\em i.e.,} they are designed to enforce irregular access.  Thus, when a
given hash table has too many unique keys to be held in on-chip
caches, loading hash table entries can become a performance
bottleneck.

\begin{lstlisting}[language=C++,label={list:genhisto},
    basicstyle=\ttfamily\scriptsize,
    numbers=left, numbersep=4pt,xleftmargin=2pt,
    caption={Example Histogram Calculation using C++ STL
      unordered\_map, illustrating memory-bound DILs.}]
#include <unordered_map>

typedef unsigned long UINT64;
typedef unordered_map<UINT64, UINT64> Histo;

void gen_histo(UINT64 *array, UINT64 size, Histo &histo)
{
    for(UINT64 i=0; i < size; i++)
        if (histo.find(array[i]) != histo.end())
            histo[array[i]]++;
        else
            histo[array[i]] = 1;
}
\end{lstlisting}

\begin{figure*}[htb]
\centering
\subfloat[Disassembly.\label{fig:histodisasm}]{\includegraphics[scale=0.70]{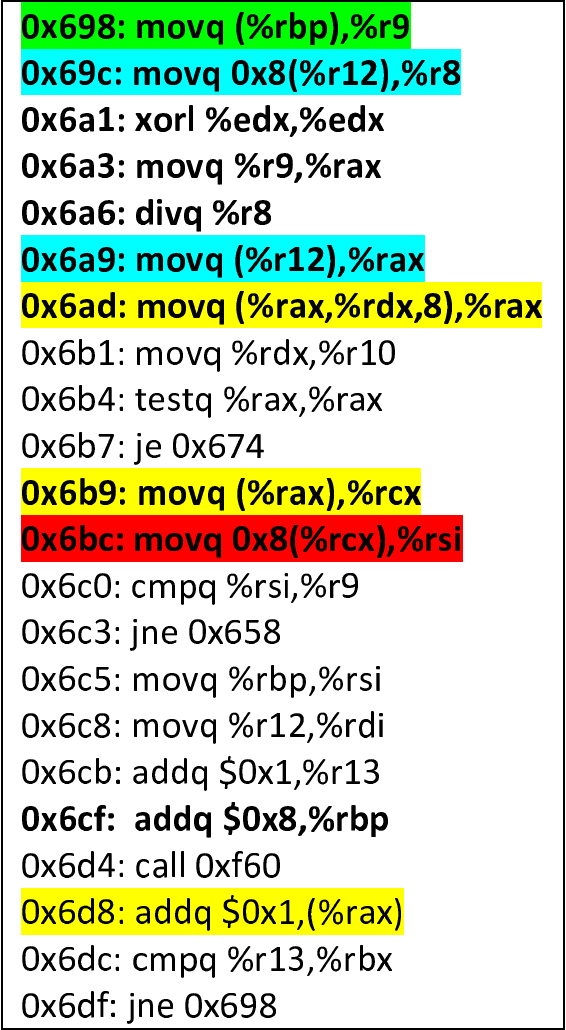}}
\hspace{0.5cm}
\subfloat[Dataflow.\label{fig:histodataflow}]{\includegraphics[width=0.60\textwidth]{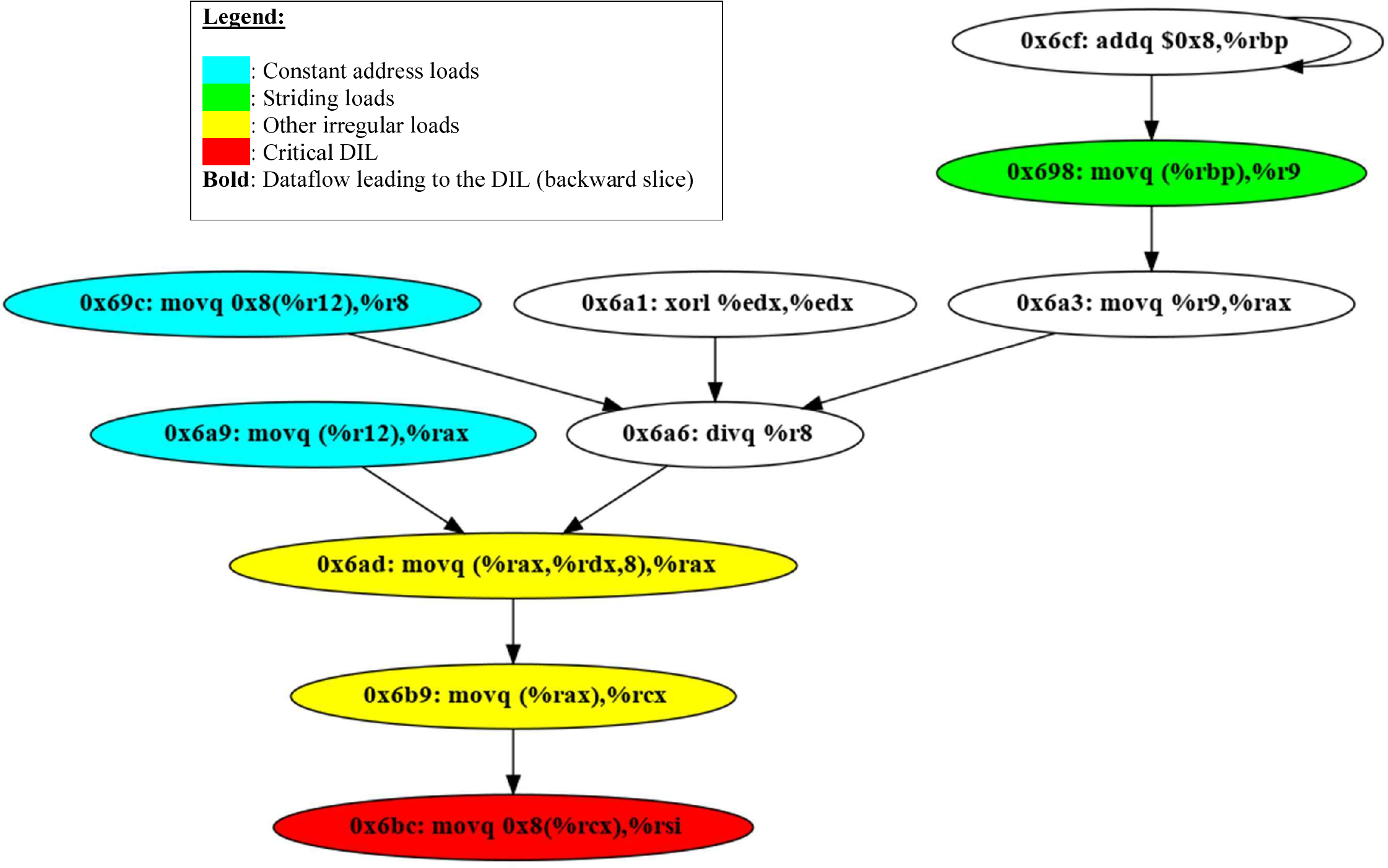}}
\caption{Disassembly and dataflow of the hot loop in Listing~\ref{list:genhisto}.}
\label{fig:histoloop}
\end{figure*}

Listing~\ref{list:genhisto} shows a sample frequency histogram
computation over an integer array that uses C++ STL unordered\_map,
the standard implementation of a hash table, to store the frequency
counts.  The code assigns a frequency of 1 to a key encountered for
the first time and increments the frequency every time the same key is
subsequently encountered.  Assuming each key is encountered several
times, the hot path around the loop is through the frequency increment
shown in Line 10. Figure~\ref{fig:histodisasm} shows the x86\_64
disassembly for this short, hot path around the loop when compiled with
\texttt{gcc 6.1} using the \texttt{-O3 -march=native} flags. Please 
note that this disassembly is part of the implementation of
unordered\_map and that longer, cold paths around the loop are not 
shown for clarity. It highlights the load instructions in the loop with
different colors: constant address loads are shown in blue,
striding loads in green, and irregular loads in yellow or
red.  Microarchitectural simulation of this loop shows that the average
cycles per instruction (CPI) measured at retirement (as the number of
cycles between the retirement of the current dynamic instruction and
the previous dynamic instruction) is the highest for the load
instruction at instruction pointer (IP) \texttt{0x6bc}. We call this
the {\em critical DIL}.  Note that the load instruction at
address \texttt{0x6d8} is also irregular but not delinquent.  This is
because it produces addresses that are small constant offsets from
those produced by \texttt{0x6bc} and hence fall into the same cache
line.

To see why it is difficult for the microarchitecture to execute this
loop, Figure~\ref{fig:histodataflow} shows the backward
slice~\cite{dynamic_slicing} of the critical DIL \texttt{0x6bc} (red: 
it is critical because it stalls the pipeline after becoming the most 
senior instruction in the re-order buffer waiting for data from memory).
The backward slice captures the dataflow between successive iterations
of the DIL.  An edge from a lower IP to a higher IP indicates the
dataflow within an iteration while an edge from a higher IP to a lower
IP indicates the dataflow from a previous iteration of the loop.  A
cycle in this graph indicates a loop-carried dependence.  We can see
that a single striding load \texttt{0x698} (green) feeds all the DILs
in this loop.  In Listing~\ref{list:genhisto}, this striding load
corresponds to the variable \texttt{array}. Hardware prefetchers 
prefetch this striding load successfully. Looking at the path from
the striding load to the critical DIL (green to red path) in the
backward slice, we can observe that as part of the hashing function,
the value from the striding load is divided by a constant and the
remainder is used to calculate the address of the DIL in successive
indirections. These indirections occurring after the non-linearity 
(due to the \texttt{div} instruction) are beyond the capability of 
hardware prefetchers today. When the number of unique keys in the hash table is too
large to fit in the on-chip caches, each of these indirections becomes
DRAM-bound.  Even with large out-of-order instruction windows, the
latency of three consecutive round-trips to DRAM becomes
impossible to hide. 

Neither IMP~\cite{imp} nor
runahead~\cite{runahead} improves the performance of this loop. The
non-linear relationship between the
value of the striding feeder load and the consumer DIL is outside the
purview of IMP, which only captures linear relationships of the form
\(ax+b\).  Runahead, on the other hand, can alleviate branch
mis-speculations, but the chain of dependent cache misses will ultimately
cause the runahead engine to stall for data.

\subsection{Helper Thread Implementation}
Before delving into our approach, we discuss the challenges in
implementing a prefetcher using traditional helper threads.  Prior
work has studied several design choices including hardware support for
extremely lightweight threads~\cite{helper_cf04} and a variety of
trigger mechanisms, including one helper thread spawning another in
series~\cite{helper2_isca01}. However, our requirement is that the
prefetcher must be able to run on existing CPUs without any additional
hardware or firmware support.  Hence, we choose the \emph{clone}
system call in Linux~\cite{clone_syscall} to create helper threads.
As a first step, we measure the overhead of spawning a thread using
this approach to be approximately 3-30 $\mu$s, which is equivalent to
several tens- to hundreds-of-thousands of CPU cycles in our test
system.  Next, we spawn the backward slice of the critical DIL as a
separate thread at each entry into the loop.  Since the backward slice
is much smaller than the main loop, it runs ahead despite the 3-30
$\mu$s delayed start.

Note that the loop has calls to functions that allocate memory
on-demand for the hash table.  Thus, if the helper thread runs
arbitrarily ahead, it can cause segmentation faults for the main
program by accessing memory that has not yet been allocated.  Here, we
make two design decisions: (1) we exit the helper thread at all
paths other than the hot path through the loop; and, (2) to balance
the overhead of thread spawning against the performance benefit due to
prefetching, we skip a tunable fraction of the loop before starting
the helper threads.  This allows time for most memory
allocation to complete before we begin prefetching.  

\begin{figure}[h]
    \centering
    \includegraphics[width=\columnwidth]{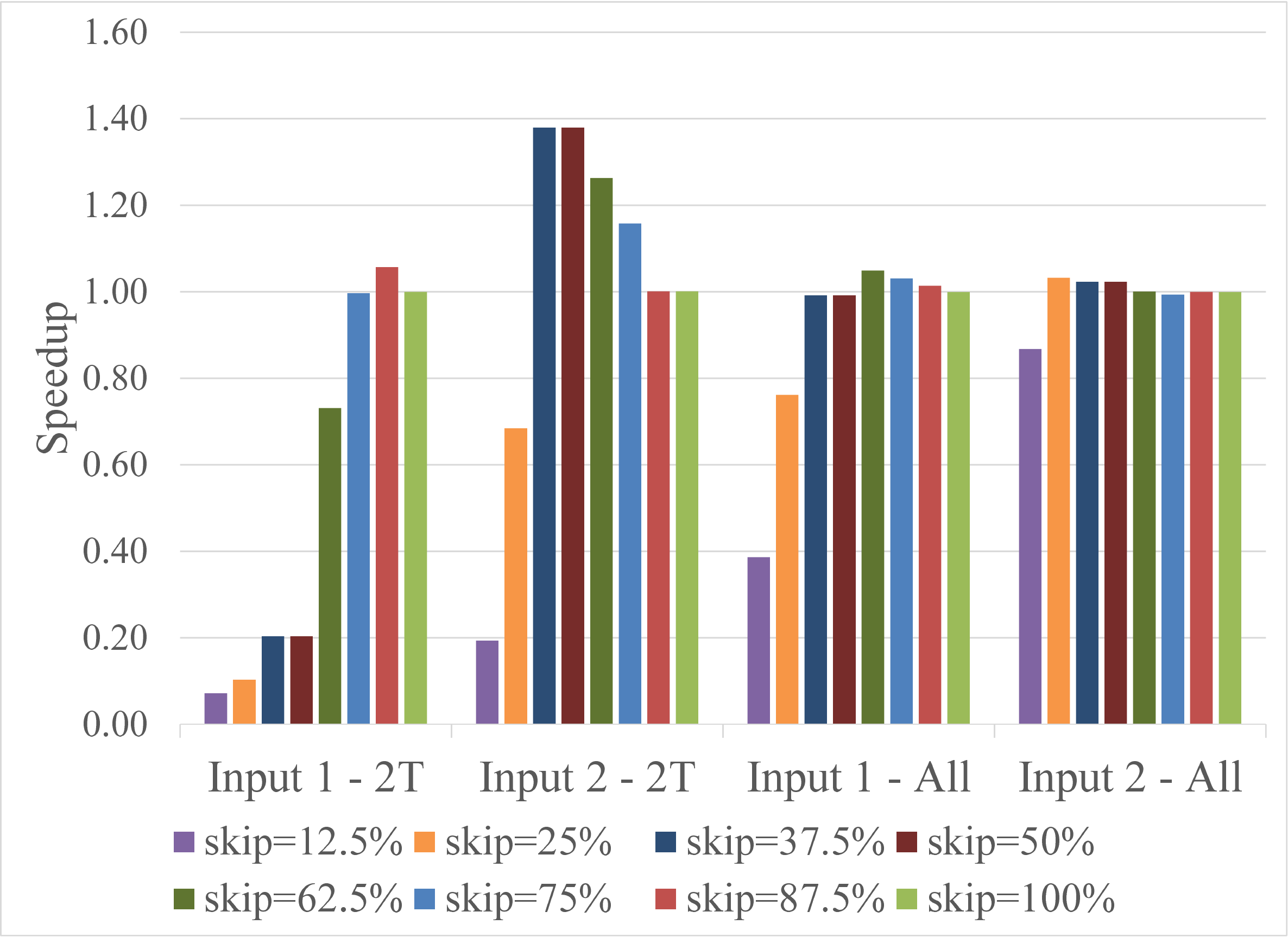}
    \caption{Performance results from the helper thread implementation. 
      The two plots on the left are from when the main and helper threads
      are restricted to SMT contexts in the same core. The two plots
      on the right are obtained by allowing the threads to be scheduled
      on any of the available cores.}
    \label{fig:helpertuning}
\end{figure}

We run the application with the helper threads for two different
inputs and examine the speedup over the baseline implementation
without prefetching.  We run the helper version in two different
modes: first, we allow the helper thread to run only on the same core
as the main thread.  In this case (2T), the main and helper threads use the
two SMT contexts of the same core. In the second mode (All), we allow the
threads to be scheduled in any of the cores available on the
machine. The speedup is shown in Figure~\ref{fig:helpertuning} for
different settings of the tunable thread start delay (skip).

When the tunable skip thresholds are low, the helper threads are
created and destroyed too often alongside the frequent memory
allocation.  When there is no restriction on the number of parallel
thread contexts available, this does not cause too much slowdown
(right, ``All''), but with only two SMT contexts on the same core, the
spawning penalty is prohibitive (left, ``2T'').  On the other hand,
high skip thresholds result in lost opportunity.  Moreover, the
maximum speedups on the right are low, due to cache interference
caused by threads hopping to different cores.  Furthermore, the
optimal skip threshold for the 2T case varies across inputs (it is
87.5\% for Input 1 and 37.5-50\% for Input 2). This serves to
illustrate how tricky it can be to tune the helper thread
implementation.

\section {Method}
\label{sec:method}

In the previous section, we explained the problem of memory-bound DILs
through a hash table example and outlined the challenges in implementing
a prefetcher with helper threads. Here, we will outline our approach to a
solution, with a reminder that we want to create a prefetcher implementation
without threads.

Observing the backward slice shown in Figure~\ref{fig:histodisasm}, we
see that the one cycle in the graph is comprised of a single
instruction \texttt{0x6cf}, \emph{i.e.}, the stride address increment,
and that it is the only loop-carried dependence in this backward
slice.  Note that a cycle in the backward slice captures the
essential relationship between the addresses produced by a DIL in
successive iterations.  If the instructions in the cycle can be
executed efficiently by the hardware, then it becomes possible to
overcome the bottlenecks {\em outside the cycle} through software
prefetching.  Conversely, if the cycle cannot be executed efficiently
by the hardware due to true data dependencies, then the performance of
such a DIL cannot be improved with software prefetching.

Specifically, if the backward slice of a DIL has no cycles with any
irregular memory operations, then such a cycle can be executed
efficiently by the hardware.  Such a cycle can be run multiple
iterations ahead of the main computation by a software prefetcher and
we describe DILs with such a backward slice as {\em runnable}.  On the
other hand, if the cycles in the backward slice have delinquent irregular memory
operations, then running a few iterations ahead gives no advantage;
the performance bottleneck would simply shift from the main
computation to the prefetcher computation instead.  This is the
classic situation of pointer chasing and we refer to such DILs as {\em
  chasing} DILs.  Short of moving the whole cycle of chasing
computation closer to memory (through techniques such as processing in
memory), not much can be done to improve such loads.

\begin{figure*}[htb]
\centering
\subfloat[A Runnable DIL.\label{fig:runnable}]{\includegraphics[width=0.40\textwidth]{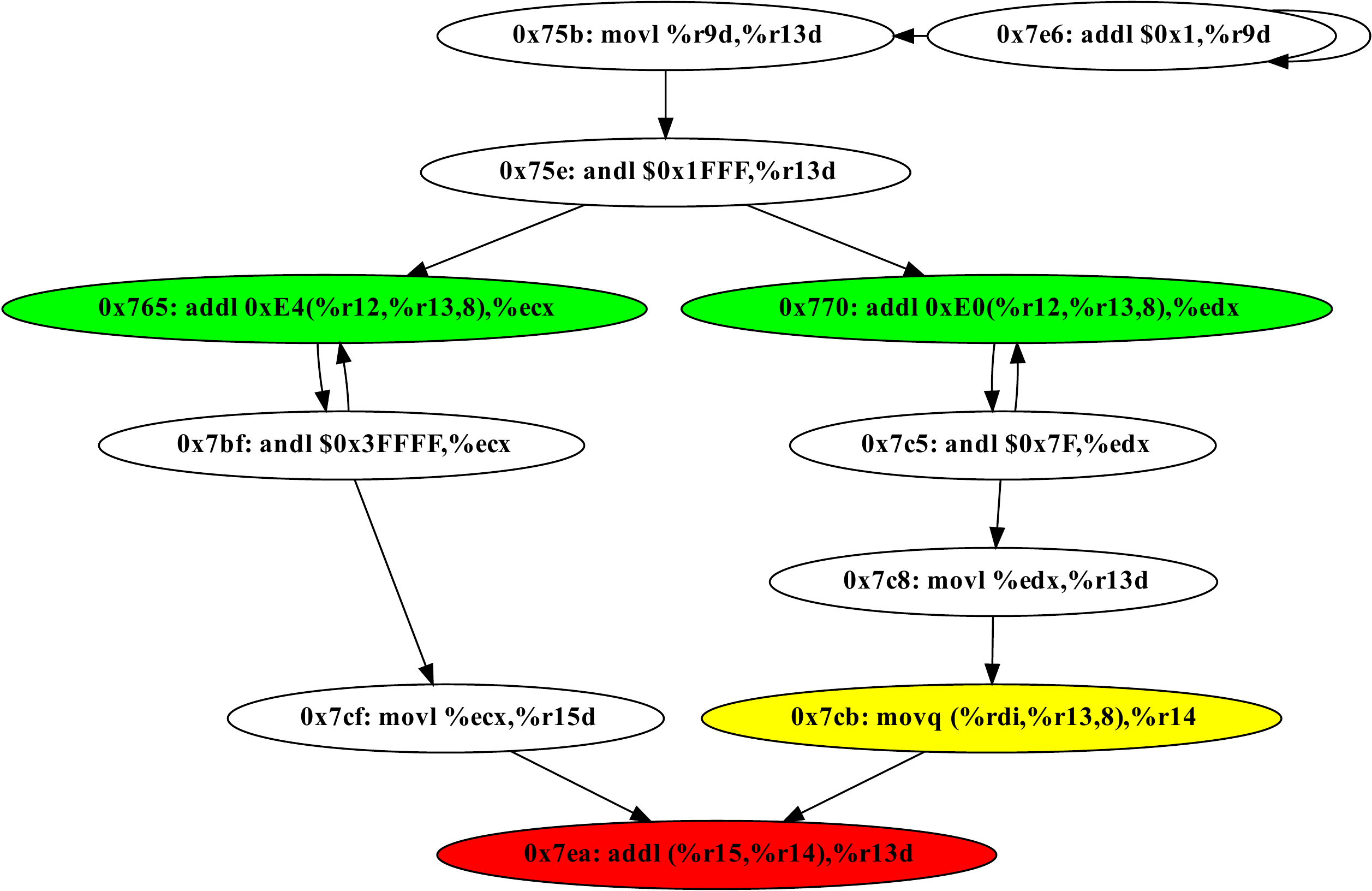}}%
\subfloat[A Chasing DIL.\label{fig:chasing}]{\includegraphics[width=0.60\textwidth]{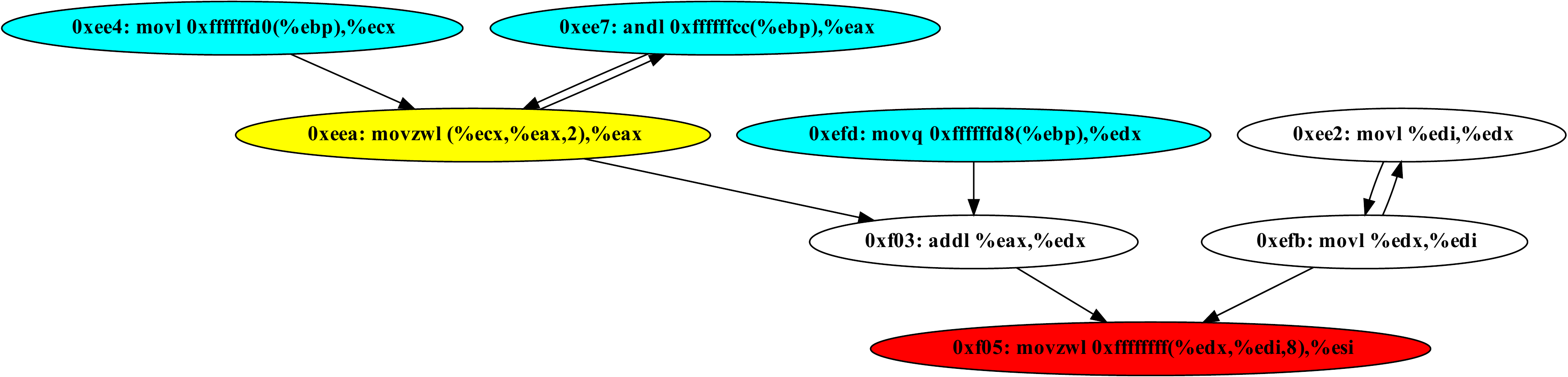}}%
\caption{Examples of a Runnable and a Chasing DIL. The runnable DIL
  has three cycles, but no irregular memory operations are part of
  these cycles.  In contrast, the chasing DIL has two cycles and one
  of them has an irregular memory operation (\texttt{0xeea}).}
\label{fig:runchase}
\end{figure*}

To explicitly contrast runnable and chasing DILs, we provide
respective examples extracted from real applications in
Figure~\ref{fig:runchase}. The coloring scheme remains the same as in
Figure~\ref{fig:histoloop}. The backward slice of
the runnable DIL shown in Figure~\ref{fig:runnable} has three total
cycles, but none have irregular memory operations.  In contrast, the
backward slice of the chasing DIL shown in Figure~\ref{fig:chasing}
has two cycles, one of which has an irregular memory operation at
\texttt{0xeea} (yellow).

Through dataflow analysis, we can determine if a DIL is runnable.
However, merely being runnable does not guarantee that a DIL is also
prefetchable.  We must also examine the control flow within the loop.
If the backward slice of a DIL varies along the different control flow
paths through the loop, then the backward slice is {\em control
  dependent} on the branches within the loop.  A popular example of
such a situation occurs in an array-based implementation of a binary
search tree.  If the current search node is at index $x$, the next
node to be searched can either be the left child (at index $2x+1$) or
the right child (at index $2x+2$), dependent on the result of the
comparison at the current node.  We exclude such scenarios by design
for two reasons: first, such situations are rare and second,
prefetcher complexity increases tremendously in such cases.  To see
why, let us consider the example of the binary tree where both the
paths are equally likely.  If we want to prefetch $k$ iterations
ahead, then there are $2^k$ possible addresses to prefetch.  We have
the option of either prefetching all of those addresses or
implementing a software-based branch predictor to select one of the
addresses to prefetch.  Both of these options are unrealistic and
hence we deliberately exclude such situations by considering only 
DILs that have backward slices that remain {\em control independent}
of all the branches within the loop.  Finally, when a DIL is {\em
  runnable} as well as {\em control independent}, we call it {\em
  prefetchable}.  These two criteria comprise our DIL screen; our
software prefetcher framework only targets DILs that pass this screen
for custom prefetching code generation.

\begin{figure}[ht!]
    \centering
    \includegraphics[width=\columnwidth]{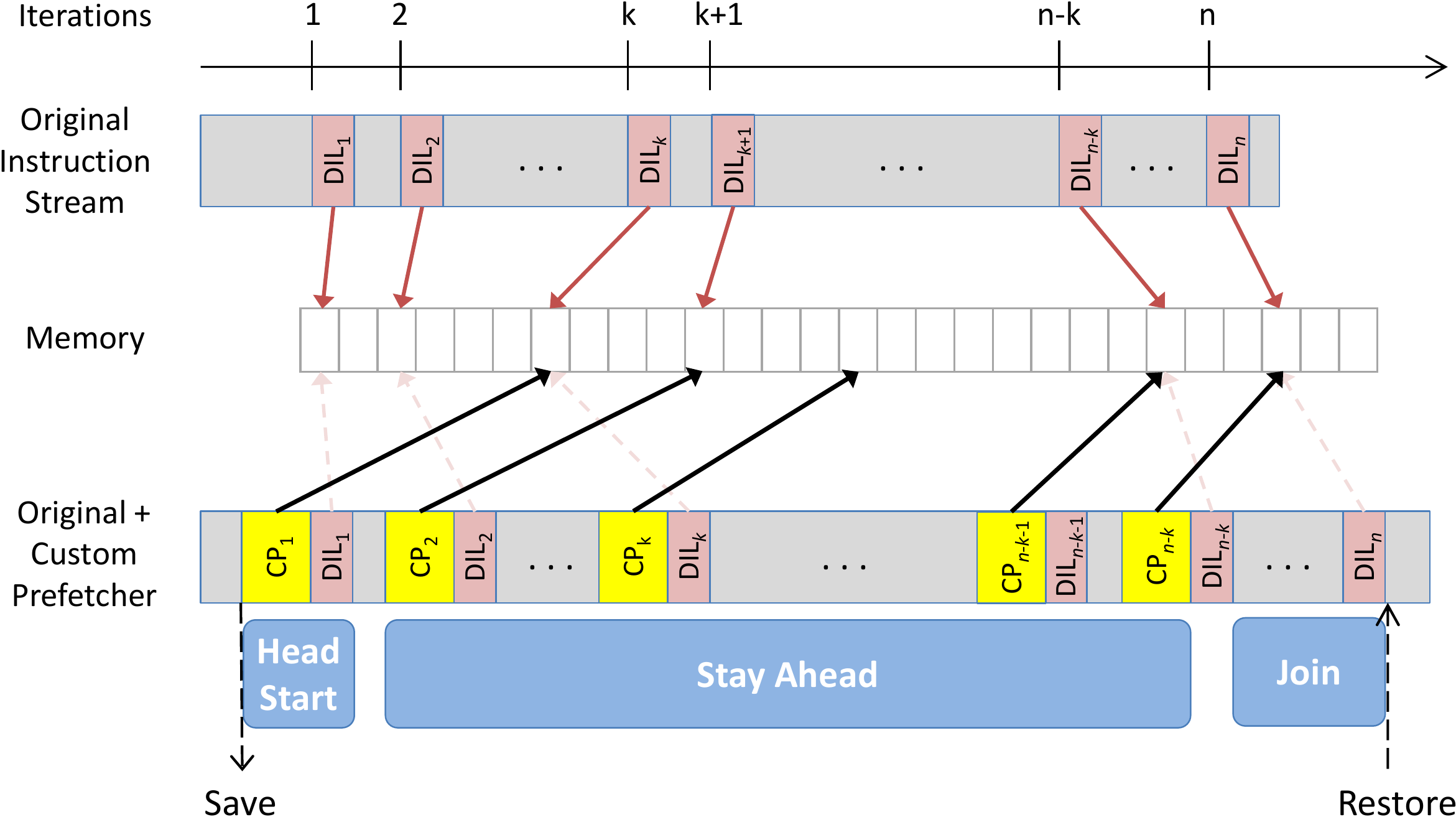}
    \caption{Overview of the phases in our prefetching scheme.  DILs
      (marked $\textrm{DIL}_i$) at each loop iteration demand particular memory addresses.
      We insert customized prefetching code (yellow) that runs $k$
      iterations ahead to prefetch those addresses and mitigate the
      delinquency.}
    \label{fig:carrothorse}
\end{figure}

Once a prefetchable DIL has been identified, inspired by software
pipelining~\cite{rau_micro81,lam_pldi88}, we take a {\em carrot and
  the horse} approach to prefetching it. We duplicate the backward
slice code and assign new registers to it.  By analogy, this code is the 
  ``carrot'' and the main computation is the ``horse''.  Prior to the
entry into the loop, the carrot is first extended $k$ iterations ahead
of the horse.  We call this phase in the dynamic execution the {\em
  head start} phase.  After the entry into the loop, the carrot locks
steps with the horse and stays a constant $k$ iterations ahead.  We
call this phase in the dynamic execution the {\em stay ahead}
phase.  During the last $k$ iterations of the loop, the carrot ceases
to stay ahead and merges with the horse.  We call this phase of dynamic
execution the {\em join} phase.  Finally, since the carrot overwrites
the architectural registers, we also need to \emph{save} them onto the stack
at loop entry and \emph{restore} them at loop exit.

This process is more formally described in
Figure~\ref{fig:carrothorse}.  The figure contrasts the dynamic
instruction streams and the memory addresses accessed before and after
the insertion of the software prefetcher code.  At the top, each
iteration in the original instruction stream has a DIL (marked $\textrm{DIL}_i$) that
demands a particular memory address.  At the bottom, customized
prefetching code (yellow) is inserted into the instruction stream.
These are given a head start to run $k$ iterations ahead such that the
addresses they prefetch mitigate all the DILs within \emph{stay ahead}
and \emph{join} phases. Please note that although the carrot and horse
approach sounds similar in principle to software pipelining, it is not a
instruction scheduling technique and the speedups are exclusively 
because of the duplication of code that stays a constant number of 
iterations ahead.

With this overall picture in mind, we provide the details of our
method next.

\subsection {Analysis and Screening}
\label{sec:analysis}

The first step is the identification of DRAM-bound load
instructions.  For this purpose, we employ detailed profiling and
dataflow analysis of the application of interest. Our analysis
infrastructure uses a pintool~\cite{pin} to generate the basic block
vector profiles of the application at a 10M instruction granularity
and the SimPoint~\cite{simpoint} methodology to identify
representative regions for microarchitectural simulation.  We
implement K-means clustering and augment it with silhouette
analysis~\cite{silhouette} to ensure clusters of good quality.  We
then use PinPlay~\cite{pinplay} to generate two traces for each
SimPoint, a {\em short} trace for functional simulation and dataflow
analysis and a {\em long} trace for cycle-accurate microarchitectural
analysis.  The short functional simulation traces are 10M instructions
long. The long microarchitectural traces are 310M instructions long in
order to accommodate in simulation a cache warmup period of 295M
instructions, microarchitectural warmup period of 5M instructions, and
a detailed cycle-accurate simulation of 10M instructions.

Next, we perform cycle-accurate simulations of a microarchitecture
resembling Intel's Skylake~\cite{skylake} CPU on an in-house x86\_64
performance simulator.  The cycle simulations produce a list of
DRAM-bound load instructions, defined as those with an average CPI
higher than the latency of the last level cache.  This output list is
sorted by the fraction of the total L1 data cache misses produced by
each load instruction.  We then select the delinquent load
instructions covering the top 99\% of all the L1 data cache misses for
further dataflow analysis.  It is worth noting that we chose this
route of implementation through trace-level, cycle-accurate
microarchitectural simulation, but there are other ways to identify
DRAM-bound load instructions, \emph{e.g.}, with assistance from hardware
performance counters~\cite{linux_perf}, functional cache
simulation~\cite{pin_analysis}, profiling DRAM accesses~\cite{concurrent_computer14,hmtt_sigmetrics08}
or even statically~\cite{panait_cgo04}.

The next step in our analysis is to identify the irregular loads from
the list of DRAM-bound loads. We achieve this through the calculation
of address deltas, defined as the numerical difference between the
addresses produced by successive executions of a load instruction. We
compute the address delta histograms for all the DRAM-bound loads in
the short traces. An n-dimensional regular array accessed inside a
loop can produce $n$ different address deltas. Hence, in order to
filter out high dimensional regular arrays common in numerical code,
we choose a threshold of 10 deltas, \emph{i.e.}, we select only those
load instructions with at least 10 distinct address deltas covering
the top 90\% of the executions. This is our DIL candidate list.

We then build the dynamic control flow graph~\cite{cfg} using the
short traces and determine the loop immediately encompassing each
DIL.  After that, we enumerate all the different control flow paths
within the loop.  For each such path, we perform dynamic dataflow
analysis~\cite{dynamic_slicing} to compute the backward slice graph
and enumerate all the simple cycles in it~\cite{graph_cycles}. With
the information from the aforementioned address delta analysis, we
find the cycles that do not involve any irregular memory operations
and determine whether the DIL is runnable.  When a runnable DIL
has the same backward slice along all the control flow paths within
its encompassing loop, we flag it as control independent and hence
prefetchable.  For these graph computations, we utilize the
networkx~\cite{networkx} package in Python.

Once all the prefetchable DILs and their encompassing loops are
identified, we group the DILs into loops and determine which among
them inside the same loop produce addresses that are at a small
constant offset from one another. We drop all such DILs from our list
except the DIL with the largest average CPI (the critical DIL) since
such addresses either fall within the same cache line as the critical
DIL or regular hardware prefetchers will handle these properly.  The
load instruction at address \texttt{0x6d8} in
Figure~\ref{fig:histoloop} is an example of such a case. 
Moreover, to avoid alias analysis, we restrict ourselves to situations 
where the addresses of the stores in the backslice can be inferred
statically. Through these successive screens, we are ultimately left 
with only those prefetchable DILs that are most challenging for the 
hardware to prefetch. 

\subsection {Prefetcher Generation}
\label{sec:codegen}

We now illustrate the generation of the customized prefetching code
for the phases shown in Figure~\ref{fig:carrothorse}, using the hash
table example from Figure~\ref{fig:histoloop}.  Keep in mind that we
do not operate on the source code and hence begin with the loop shown
in Figure~\ref{fig:histodisasm}.  We insert the prefetcher assembly
into the application's assembly directly.

As a first step, we attempt to find unused architectural registers 
inside the loop.  When there are no unused registers available, we
create new local variables on the stack and select registers to spill
onto them in the following order for minimal performance impact:
\begin {enumerate}
\item Registers only written to but never read from inside the loop 
  (only the last write to these registers need to be made visible
  outside the loop);
\item Registers only read from but never written to inside the loop
  (all references to these registers will be replaced by their
  corresponding stack loads).
\end{enumerate}

For our example in Figure~\ref{fig:histodisasm}, it turns out that
registers $r11$, $r14$ and $r15$ are unused inside the loop.  Among
these, $r11$ is caller-saved and there is a function call inside the
loop, meaning it could potentially be used inside the function
call. Thus, we choose $r14$ and $r15$ as the registers to use for our
{\em carrot} computation {\em i.e.,} inside the customized
prefetching code. As discussed before and shown in
Figure~\ref{fig:carrothorse}, the first step is to save these
registers onto the stack:

\begin{lstlisting}[language=myasm,
    basicstyle=\ttfamily\scriptsize,
    numbers=left, numbersep=4pt,xleftmargin=2pt,
    label={list:save},caption={The save phase.}]
# save unused registers at loop entry
	pushq %r14
	pushq %r15
\end{lstlisting}

Next is the head start phase, also performed at loop entry, where
the prefetch computation gets a $k$-iteration head start. In our
example, $rbp$ is the only register written inside the cycle in the
backward slice graph.  Hence, we duplicate it onto $r14$.  We also use
$r15$ as scratchpad to perform the loop boundary check by comparing it
with the loop limit in $rbx$, as follows.

\begin{lstlisting}[language=myasm,
    basicstyle=\ttfamily\scriptsize,
    numbers=left, numbersep=4pt,xleftmargin=2pt,
    label={list:headstart},caption={The head start phase.}]
# prefetch distance
	.set k, 8
# powers of 2 for easy mul
	.set logk, 3
# r14 is the carrot reg that stays ahead of the horse reg rbp
	movq %rbp, %r14
# bounds check: use r15 as scratchpad
	movq $k, %r15
	addq $0x1, %r15
	cmpq %rbx, %r15
# don't start ahead if bounds check fails
	 jge SKIP1
# stride
	movq $0x8, %r15
# k*stride
	shlq $logk, %r15
# carrot reg = horse reg + k*stride
	addq %r15, %r14
SKIP1:
\end{lstlisting}

The next two phases are the (1) stay ahead phase, where the prefetcher
(carrot) computation stays ahead of and in lock step with the main
(horse) computation, and (2) the join phase, where the prefetcher
computation no longer stays ahead and ultimately merges with the main
computation.  Both of these phases are inserted into the loop body and
are shown in Listing~\ref{list:stayjoin}.  For clarity, we distinguish
the inserted code from the existing code by highlighting the inserted
code in yellow.

\begin{lstlisting}[language=myasm,
    basicstyle=\ttfamily\scriptsize,
    numbers=left, numbersep=4pt,xleftmargin=2pt,
    label={list:stayjoin},caption={The stay ahead and join phases.},linebackgroundcolor=\bgndcolor{\value{lstnumber}}]
START:
# duplicate line 4
	movq (%r14),%r15
	movq (%rbp),%r9     
	movq 0x8(%r12),%r8  
# duplicate lines 11-13, write output to r15
	xorl %edx,%edx
	movq %r15,%rax
	divq %r8
	movq %rdx, %r15
	xorl %edx,%edx
	movq %r9,%rax
	divq %r8
	movq (%r12),%rax
# duplicate line 17
	movq (%rax,%r15,8),%r15
	movq (%rax,%rdx,8),%rax
	movq %rdx,%r10
	testq %rax,%rax
	je LABEL1
# duplicate lines 19-20, 28-29
	testq %r15, %r15
	je SKIP2
	movq (%r15),%r15
# prefetch DIL	
	prefetcht0 0x8(%r15)
SKIP2:
	movq (%rax),%rcx
	movq 0x8(%rcx),%rsi
	cmpq %rsi,%r9
	jne LABEL2
	movq %rbp,%rsi
	movq %r12,%rdi
	addq $0x1,%r13
	addq $0x8,%rbp
# duplicate line 35
	addq $0x8, %r14
# skip staying ahead and merge 
# for the last k iterations
	movq %rbx, %r15
	subq $k, %r15
	cmpq %r15, %r13
	jl SKIP3
	movq %rbp, %r14
SKIP3:
	call 0xf60
	addq $0x1,(%rax)
	cmpq %r13,%rbx
	jne START
\end{lstlisting}

The last step is to restore the saved registers at all exit points of the loop.

\begin{lstlisting}[language=myasm,
    basicstyle=\ttfamily\scriptsize,
    numbers=left, numbersep=4pt,xleftmargin=2pt,
    label={list:restore},caption={The restore phase.}]
# restore saved registers at all loop exits
	popq %r15
	popq %r14
\end{lstlisting}

After the insertion of the prefetcher code, to ensure correctness,
we compare the output of the optimized version to that
of the unoptimized version and require that they match exactly, except
for those outputs dependent on operating system behavior such as
timing measurement, random number generation, signal handling, {\em
  etc}.

\section {Experimental Evaluation}
\label{sec:eval}
Recall from figure~\ref{fig:tracedata} that while DIL prefetching 
may not benefit all applications, some irregular applications can
benefit a lot (right side of figure~\ref{fig:tracedata}). For instance, 
several high value cloud applications fall into this category. Hence, we evaluate our proposal on a set 
of irregular memory workloads similar to the work by Ainsworth and 
Jones~\cite{indirect_asplos18} (we do not use the applications from
figure~\ref{fig:tracedata} since we don't have access to their binaries)

We study three applications from their work that are bottlenecked by
DRAM-bound DILs and add two more to the 
evaluation including the hash table example we
discussed in Sections~\ref{sec:example} and~\ref{sec:method}. 
Since our focus is on single
thread performance, we utilize the serial versions of the benchmarks
for experimentation.  We compile all benchmarks with \texttt{gcc 6.1.0}
using the flags \texttt{-O3 -march=native} on an Intel Xeon E5 server
CPU.  We run all the analysis tools for prefetchable DIL
identification and generate the customized prefetching code on the
same server as well.

\subsection {Benchmarks}
\label{sec:bench}

We now provide a brief overview of the applications studied. 

\begin{itemize}
\item \textbf{STLHistogram} is the example we discussed in Sections~
  \ref{sec:example} and~\ref{sec:method}.  It generates a random array
  of integers and computes the frequency histogram of the array using
  C++ STL unordered\_map. It takes the size of the array and the
  number of unique elements in it as arguments. Microarchitectural
  performance of this application suffers when neither the input
  array, nor the frequency histogram fits inside the on-chip
  caches. We choose this benchmark due to the popularity of hash
  tables in programs and the potential for customized prefetching to
  improve performance. Please note that since open address hash tables
  are popular, we also studied a policy based implementation of
  STLHistogram. While the baseline performance of this new version was
  7X better than the unordered\_map version, the performance improvement
  opportunity was very similar to the unordered\_map version with a single
  prefetchable memory bound DIL causing most of the stalls. Hence, we
  report results only for the unordered\_map version.
\item \textbf{PageRank} is an implementation of the popular web-page
  relevance ranking algorithm~\cite{pagerank} using the C++ Boost
  Graph Library~\cite{BGL} (BGL). It is a graph algorithm that ranks a
  website based upon the ranks of the websites that link to it.

\item \textbf{HashJoin}~\cite{hashjoin} from the University of
  Wisconsin implements the join operation of a relational
  database~\cite{dbmsbook} in main memory using hash tables. The join
  operation is very common in Structured Query Language (SQL) queries.

\item \textbf{Graph500CSR} is part of the Graph500~\cite{graph500}
  benchmark suite designed to rate supercomputer systems on their
  data-intensive performance. It performs Breadth-First Search (BFS)
  on a large graph implemented using a compressed sparse rows (CSR) data structure.

\item \textbf{Cuckoo}~\cite{halo_isca19} is an application modeling 
packet processing in the context of Network Function Virtualization 
(NFV) using the cuckoo hashing algorithm~\cite{cuckoo_algo04}.

\end{itemize} 

We run the sequential versions of these applications on the inputs
shown in Table~\ref{table:input1} and generate traces as discussed in
Section~\ref{sec:analysis}. An automatic tool analyzes the traces to produce the list
of prefetchable DILs, the loops they belong to, and a list of available
registers for code generation.  The customized prefetching code is then generated
semi-automatically with manual intervention. Specifically, our scripts generate a 
skeletal prefetcher code with the duplicated backslice and a list of candidate registers. 
However, register fills/spills, null-pointer skips and handling slices across function 
calls are done manually, Another automatic tool then statically rewrites 
the original function in the binary with a dynamic version that allocates the optimized 
code in the heap and calls it through a function pointer. We then run the optimized
binary to ensure that its output matches the original.  For
performance measurement, we employ an Intel Core i9-7900X Skylake CPU
with all the hardware prefetchers enabled, running at 3.3 GHz and
frequency scaling disabled in the BIOS. We choose an evaluation system
that is different from the one used for compilation to simulate a binary-only
scenario. We run the applications five times 
each and record the median wall clock time
before and after optimization.  We also measure the dynamic instruction
overhead of the optimized versions using a pintool~\cite{pin}. The
last column of Table~\ref{table:input1} shows the dynamic
instruction counts of the main computation in the original
applications.

\begin{table}[htb]
  \centering\small
  \begin{tabular}{|p{1.0in}|p{1.1in}|p{0.7in}|}
    \hline
    \textbf{Benchmark} & \textbf{Input} & \textbf{Dynamic Instr (B)}\\
    \hline
    \hline
    STLHistogram & 100M array, 10M unique elements & 7.9\\
    \hline
    PageRank~\cite{BGL,pagerank} & web-Google.txt~\cite{snap_google} & 1.1\\
    \hline
    HashJoin~\cite{hashjoin} & 016M\_build.tbl, 256M\_probe.tbl & 55.8\\
    \hline
    Graph500CSR~\cite{graph500}& -s 18 -e 10 & 11.2\\
    \hline
	Cuckoo~\cite{halo_isca19,cuckoo_algo04} & 8M flows & 10.2\\
    \hline
  \end{tabular}
  \caption{Benchmarks and inputs (Input 1).}
  \label{table:input1}
\end{table}
 
\subsection {Results and Discussion}
\label{sec:results}

\subsubsection {Results of Profiling and Analysis}

First, we present the results of the control and dataflow analyses for
the applications.

\begin{table}[htb]
  \centering\small
  \begin{tabular}{|p{0.75in}|p{0.2in}|p{0.5in}|p{0.3in}|p{0.8in}|}
    \hline
    \textbf{Benchmark} & \textbf{DILs} & \textbf{Prefetch-able DILs} & \textbf{Loops} & \textbf{Function Name}\\
    \hline
    \hline
    STLHistogram & 4 & 3 & 1 & gen\_histo\\
    \hline
    PageRank & 4 & 4 & 2 & pagerank\\
    \hline
    HashJoin & 2 & 2 & 1 & realprobeCursor\\
    \hline
    Graph500CSR & 6 & 6 & 2 & make\_bfs\_tree \newline verify\_bfs\_tree\\
    \hline
    Cuckoo & 3 & 2 & 1 & rte\_hash\_lookup\newline\_bulk\_data\\
    \hline
  \end{tabular}
  \caption{Results of control and dataflow analyses.}
  \label{table:dataflow}
\end{table}

The data in Table~\ref{table:dataflow} shows that of the 19 total
DILs, 17 are prefetchable. We proceed with the performance evaluation 
of the prefetchers for these DILs.

\subsubsection {Prefetcher Performance}

\begin{figure*}[htb]
\centering
\subfloat[Speedup.\label{fig:speedup1}]{\includegraphics[width=0.50\textwidth]{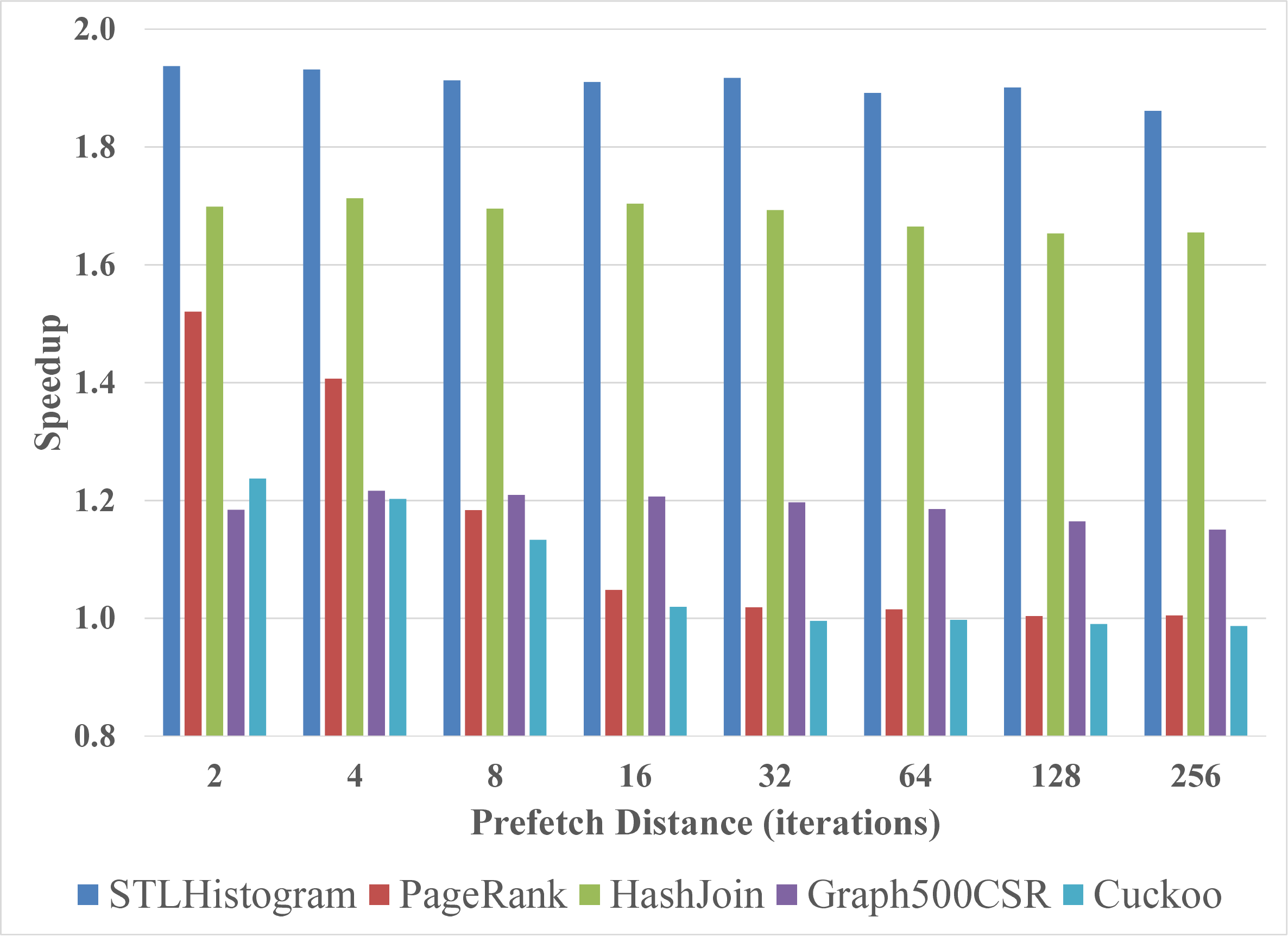}}%
\subfloat[Dynamic Instruction Overhead.\label{fig:overhead1}]{\includegraphics[width=0.50\textwidth]{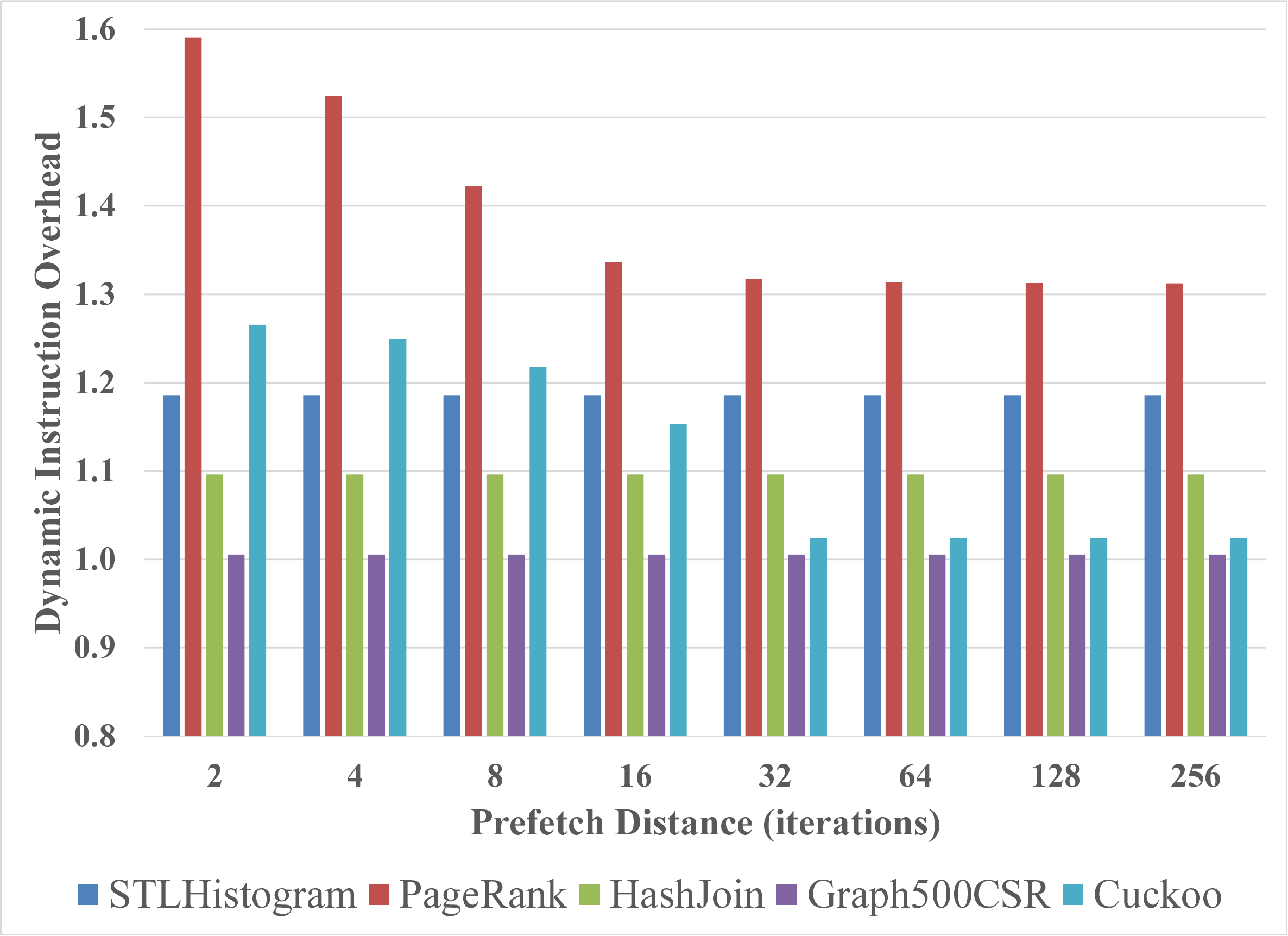}}%
\caption{Performance of the prefetching scheme (Input 1).}
\label{fig:perf1}
\end{figure*}

For the five applications, we vary the prefetch distance from
two iterations to 256 iterations in powers of two.  Note that we choose
powers of two for only a minor convenience in code generation since
multiplication can be replaced with shifts; it is not a fundamental
restriction in our approach and can easily be changed to accommodate
any arbitrary lookahead.  We then verify that the outputs of the
optimized binaries match with the original ones and then measure the
performance of the optimized versions. The speedup from the
performance optimization is shown in Figure~\ref{fig:speedup1}. The
corresponding dynamic instruction overhead is shown in
Figure~\ref{fig:overhead1}.  The $x$-axis on both the figures is the
prefetch distance, which is the number of iterations of lookahead
available for the prefetcher. On the $y$-axis in
Figure~\ref{fig:speedup1} is the ratio of the median wall clock time
before optimization to that after. Figure~\ref{fig:overhead1} plots on
its $y$-axis the ratio of the total dynamic instructions of the
baseline to that of the the optimized executions.  \emph{Note that the
speedups reported include dynamic instruction overhead since
we measure wall clock time.}

For the applications and inputs described in Table~\ref{table:input1},
there is a significant speedup of 21-94\% due to our software
prefetchers.  This speedup is in spite of significant dynamic
instruction overhead in some cases.  Hence, this result clearly
demonstrates that we are successful in accurately prefetching the
critical load addresses in a manner that does not interfere with the
memory bandwidth or with any hardware prefetchers.

A pattern to observe in the data is that even with only a few
iterations of the prefetch distance lookahead, the performance
increases significantly.  In fact, except for PageRank and Cuckoo, the
performance improvement is stable across the entire range of prefetch
distances.  This is because the loop sizes are such that only a few
iterations fit in the dynamic instruction window of the CPU. Hence, 
even with a small lookahead, the prefetcher reaches outside the instruction
window to be effective. However, the behavior of PageRank and Cuckoo deserve 
further explanation.

PageRank operates on the Web-Google graph dataset~\cite{snap_google}, which
has an average degree of less than six. The inner loop encompassing
the prefetchable DIL iterates over all the neighbors of a graph node. Hence,
the trip count of this loop is equal to the average number of a node's
neighbor or its average degree.  Therefore, prefetch distances longer
than six skip the loop fully and do not help much. This behavior can
also be seen in Figure~\ref{fig:overhead1} in the dynamic
instruction overhead data. A similar behavior occurs in Cuckoo as well, 
where the prefetchable DILs are from an inner loop with a fixed iteration 
count of 32. The lost opportunity cost due to small iteration counts
is the reason for the reduction in performance with increasing lookahead.
 
Note that the performance gains for STLHistogram and
HashJoin are much higher than those for the remaining three. In
the former two, the critical DIL is fed by a strided load
after passing through a hash function and multiple
indirections.  However, in the latter three, the strided
load feeds the DIL directly through fewer indirections (and a hash 
function in Cuckoo).  Thus, as
discussed in Sections~\ref{sec:example} and~\ref{sec:method}, the
bottleneck of the chain of dependent cache misses is much larger for
the former applications than the latter.  Consequently, the performance
boost obtained by mitigating them is also higher.

\paragraph{Impact of Inputs}

\begin{figure*}[htb]
\centering
\subfloat[Speedup.\label{fig:speedup2}]{\includegraphics[width=0.50\textwidth]{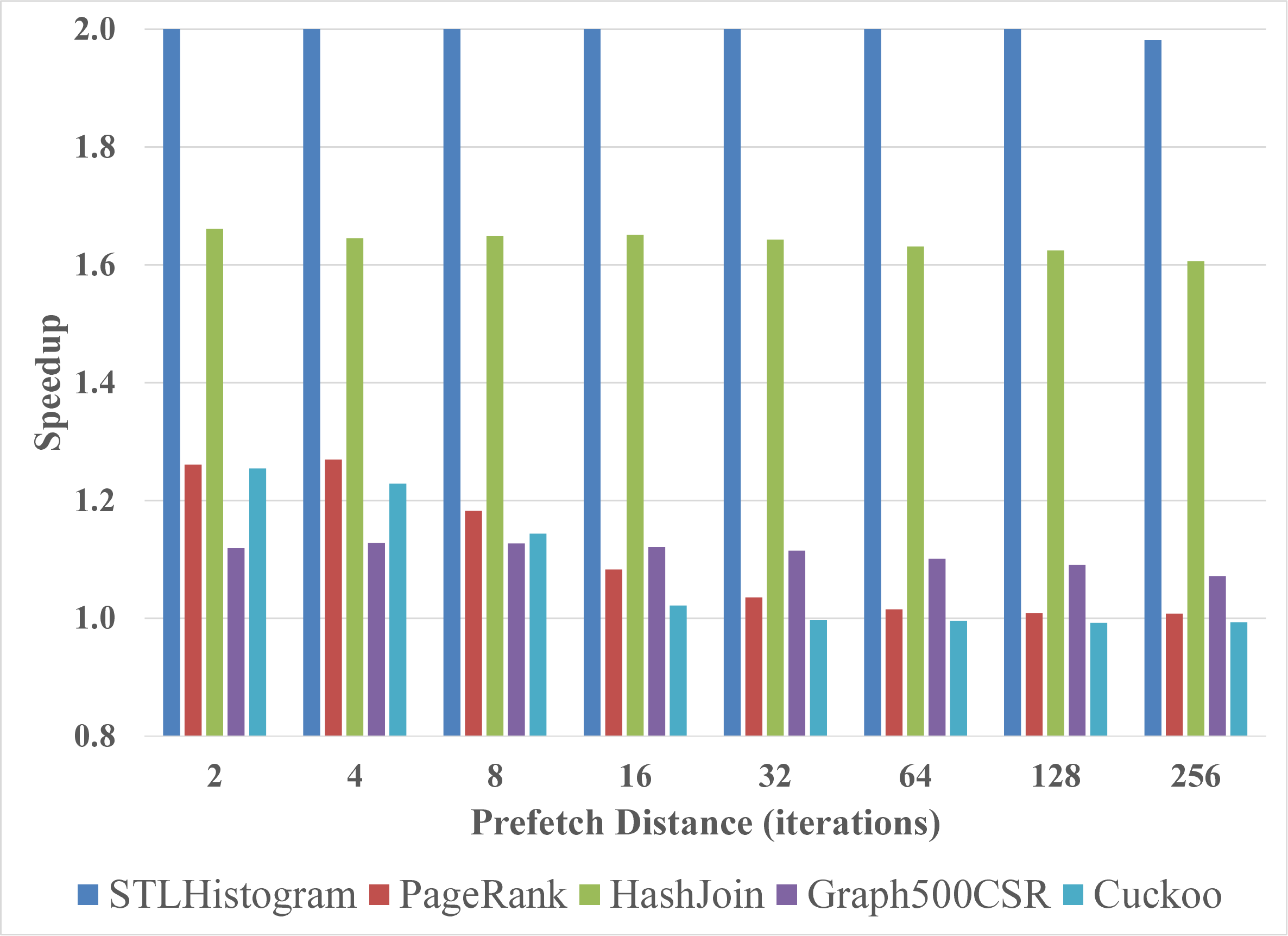}}%
\subfloat[Dynamic Instruction Overhead.\label{fig:overhead2}]{\includegraphics[width=0.50\textwidth]{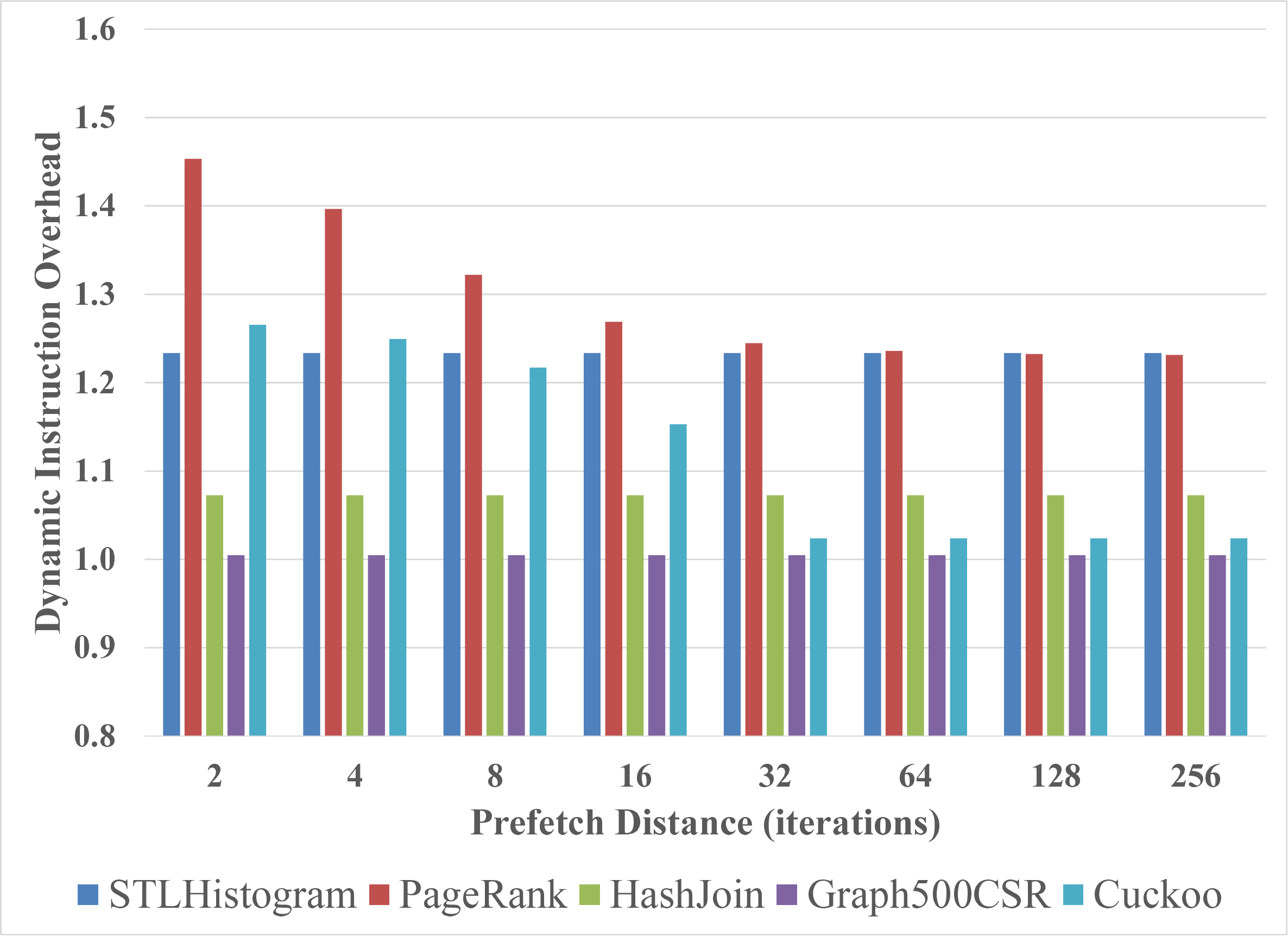}}%
\caption{Prefetcher performance on different input data (Input 2).}
\label{fig:perf2}
\end{figure*}

Next, we select a set of larger inputs for our applications and run
the optimized binaries on this set to study sensitivity to different
application inputs. Table~\ref{table:input2} lists the new inputs used
for this experiment.

\begin{table}[htb]
  \centering\small
  \begin{tabular}{|p{1.1in}|p{1.1in}|p{0.6in}|}
    \hline
    \textbf{Benchmark} & \textbf{Input} & \textbf{Dynamic Instr (B)}\\
    \hline
    \hline
    STLHistogram & 200M array, 10M unique elements & 12.7\\
    \hline
    PageRank & cit-Patents.txt~\cite{snap_patents} & 4.2\\
    \hline
    HashJoin & 032M\_build.tbl, 512M\_probe.tbl & 148.2\\
    \hline
    Graph500CSR & -s 21 -e 10 & 90.7\\
    \hline
	Cuckoo & 16M flows & 20.5\\
    \hline
  \end{tabular}
  \caption{Alternative inputs for the optimized benchmarks (Input 2).}
  \label{table:input2}
\end{table}

Figure \ref{fig:perf2} displays the speedup and the dynamic
instruction overhead for the optimized binaries running on these new
inputs.  We can see that the speedup has improved for STLHistogram,
stayed about the same for HashJoin/Cuckoo and decreased for PageRank and
Graph500CSR.  Overall, the speedups range from 10\%-100\% and are
still significant over the baselines.  For PageRank, the cit-Patents
dataset~\cite{snap_patents} has an average degree of $4.4$ which is
less than the previous Web-Google dataset.  Thus, as discussed
earlier, the drop in its speedup can be attributed to the reduced trip
count of its inner loop.  As for Graph500CSR, the new input has a
higher number of vertices but the same average degree as before and
the performance contribution of the DILs is lower than before. Hence, 
the corresponding speedup by prefetching them is also lower.

\paragraph{Impact of Microarchitecture}

\begin{figure}[h]
    \centering
    \includegraphics[width=\columnwidth]{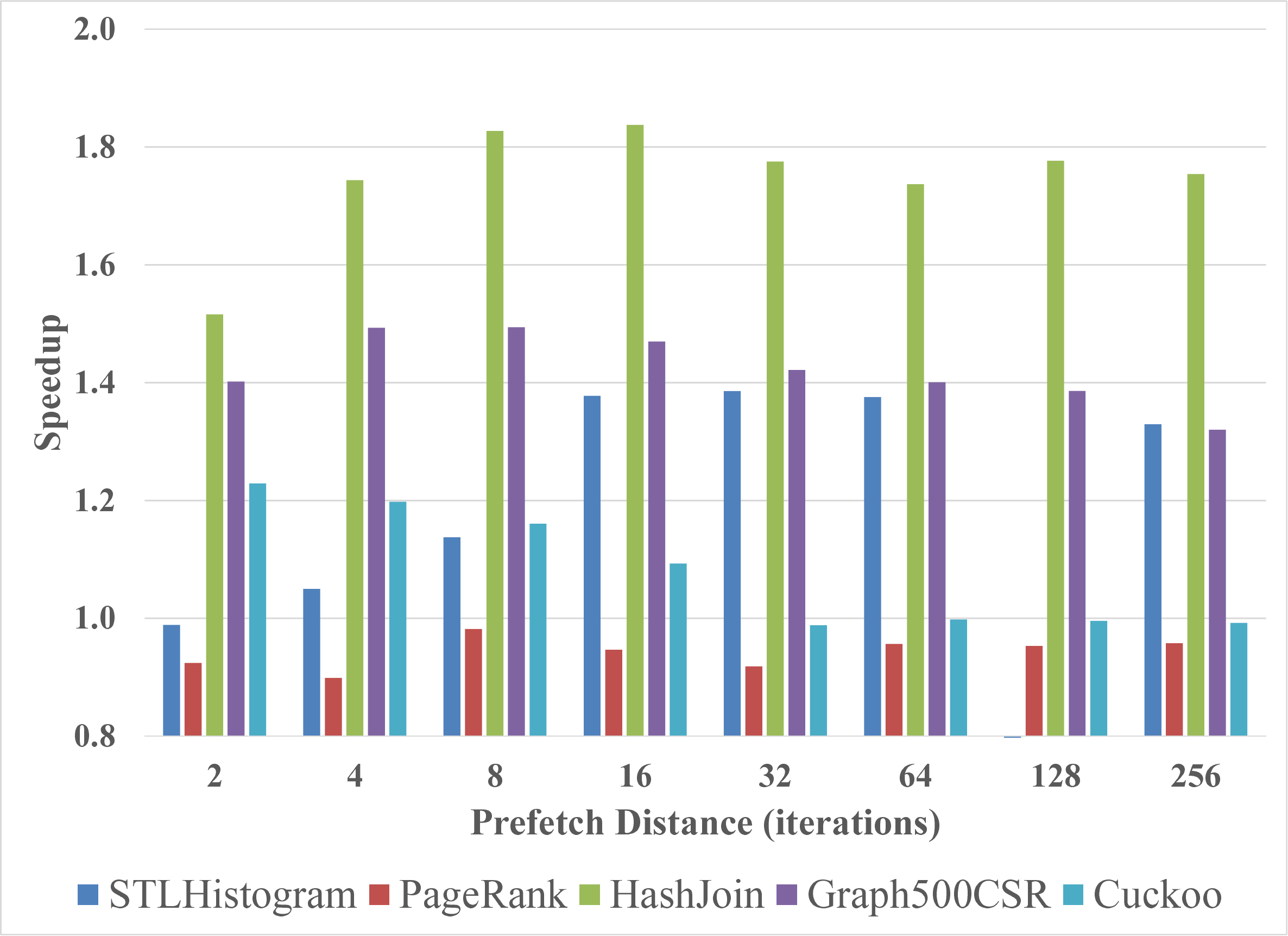}
    \caption{Prefetcher performance on a hypothetical microarchitecture
	that is 2X wider, 3.5X deeper and includes aggressive prefetchers
	similar to VLDP~\cite{vldp} and IMP~\cite{imp}}
    \label{fig:speedup3}
\end{figure}

The results shown so far were for a single microarchitecture. To study 
the impact of a different microarchitecture, we generate traces from
the unoptimized and optimized binaries and perform cycle accurate simulations 
on them for an aggressive hypothetical microarchitecture that is 2X wider and 3.5X deeper 
than Skylake. We also model two aggressive hardware prefetchers similar to VLDP~\cite{vldp}
and IMP~\cite{imp} since they were published after the release of the Skylake microarchitecture. 
Figure \ref{fig:speedup3} shows the result of the experiment.
Unlike Skylake, the dynamic instruction window of the hypothetical microarchitecture
can hold many more iterations of the loops. Hence, short prefetch distances 
do not go beyond the instruction window. This is why the speedups are lower for
shorter lookaheads (for the benchmarks without small loop iteration counts). However, 
once the prefetch distances are sufficient to look beyond the instruction window, 
the speedups stabilize afterwards. The extra latency hiding offered by the 3X increase 
in out-of-order depth causes the DILs from Pagerank to not be the bottlenecks of
performance anymore. Hence, the instruction overhead for the benchmark shows 
up as a slowdown in the chart. Nevertheless, the speedups continue to be
signifcant overall(the bar for STLHistogram is missing for the prefetch distance
of 128 due to simulation failure). 

The stability of speedups across prefetch distances beyond a particular threshold 
is helpful in case of variable DRAM latencies. Setting the lookahead for the 
worst-case memory latency can provide speedups that are robust to the variability.
Moreover, the fact that the speedups remain significant even under contemporary 
aggressive hardware prefetchers, emphasizes that our approach is complementary to 
hardware and minimizes interference.

\paragraph{Comparison with Helper Threads}

\begin{figure}[h]
    \centering
    \includegraphics[width=\columnwidth]{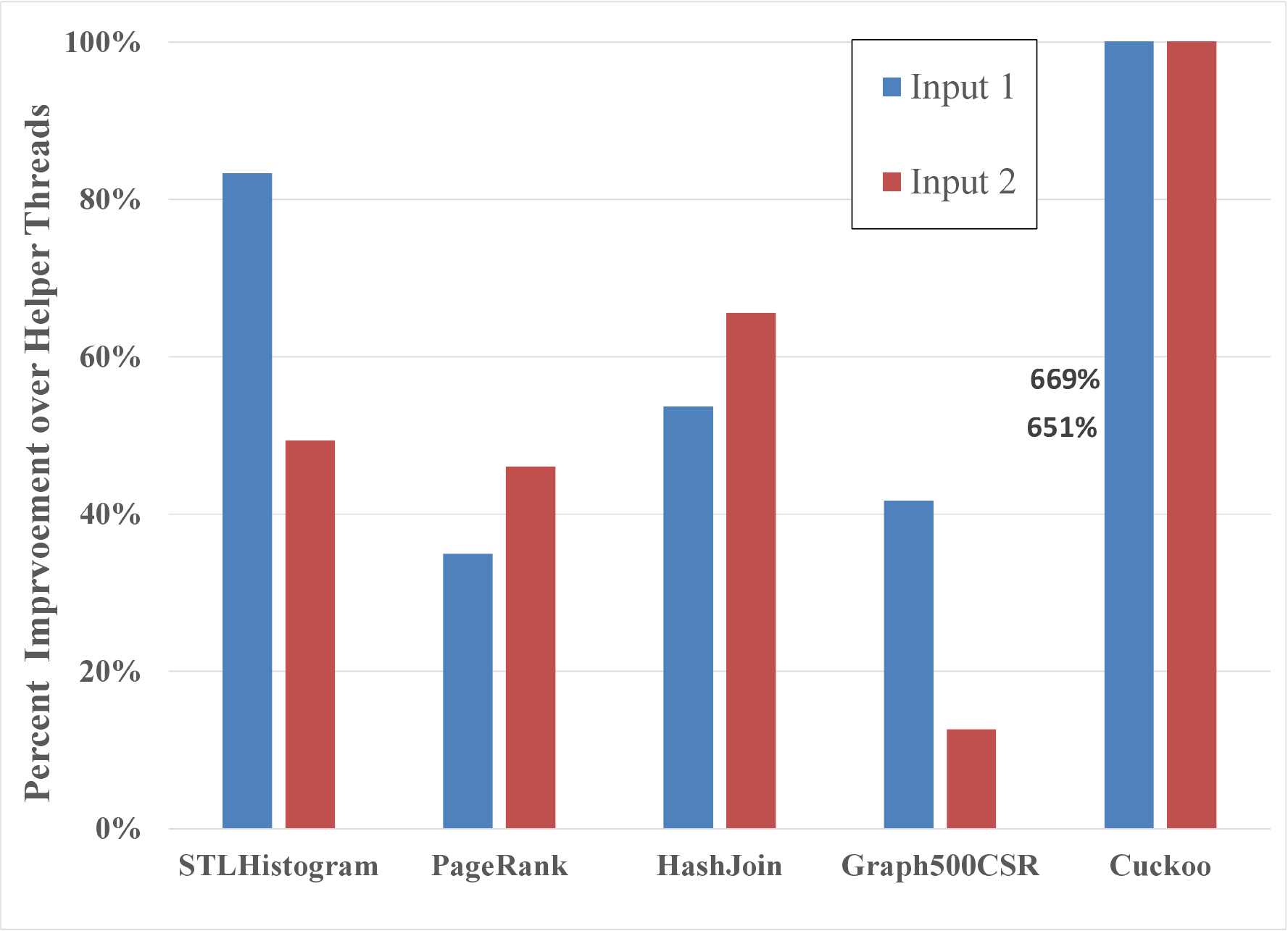}
	\caption{Percent improvement of inline prefetcher over helper
          threads.}
    \label{fig:helpercompare}
\end{figure}

We now compare the inline prefetcher to traditional helper threads. To provide the techniques 
with equal hardware, we restrict the helper thread implementations to one additional SMT 
context from the same core as the main thread. We also select the best tuning parameters
(prefetch distance for the inline prefetcher and launch trigger/frequency for helper threads)
for both the schemes. Figure \ref{fig:helpercompare} shows the results of the experiment.

Our inline prefetcher outperforms helper threads due to the latter's
thread spawning overhead. Dropping the outlier (Cuckoo), the speedups range from 
13-83\%, which is significant. For Cuckoo, the number of thread spawns is prohibitive
for helper threading to be competitive. From the results of these experiments, we conclude that the proposed
prefetcher scheme is accurate in targeting the critical load
instructions and improves single thread performance of the targeted
applications significantly. It does so without the requirements 
of traditional helper threading such as idle thread contexts and
special support from hardware or firmware. 

\subsubsection{Limitations and Future Work}

As a binary modification technique, debuggability 
can be affected due to optimization. Hence, it is a good idea 
to restrict optimization only to performance critical code. 

As a prefetching scheme running on the CPU, we drop all the
pointer chasing loads from the purview of our optimization. Such a
restriction is not essential. The backward slices and cycles with
chasing loads are ideal for offload into Processing-In-Memory
(PIM). Future work could explore means of implementing such
offloading. Also, we have restricted ourselves to software 
implementation on existing hardware, which is not mandatory. The
profile-based, offline dataflow analysis could advice 
hardware-software co-design and prefetchers could be implemented
in custom hardware instead. With the advent of Field Programmable 
Gate Arrays (FPGA), custom hardware prefetchers closely coupled 
with a processor’s pipeline are another potential direction of 
investigation.

\section {Conclusion}
\label{sec:end}

In this paper, we have described an inline software prefetcher for
DRAM-bound Delinquent Irregular Loads (DILs). In order to avoid
interfering with the hardware prefetchers for regular loads and to
keep the bandwidth impact and cache pollution to a minimum, we have
designed the scheme to be highly selective in targeting only the DILs
most difficult for the hardware to prefetch. In spite of being
selective, our approach has a significant potential for performance
enhancement as demonstrated by four applications from different
domains: a C++ hash table implementation, the PageRank website ranking
algorithm, a database join algorithm and the Graph500 breadth-first
search of a large graph. Across all inputs to the test applications,
speedup due to our inline prefetchers ranged from 10\% to 100\% on a
high-end Intel Skylake system.

Our approach performs better than a traditional implementation of
helper threads due to the latter's thread spawning overhead.  It does
so while still not requiring separate thread contexts or special
hardware/firmware support.  It makes the implementation and debugging
of the helper easier since it avoids explicit synchronization and
stays a constant number of iterations ahead of the main computation,
As a software approach that does not require high level source code,
it can be attractive for third party cloud applications.




\bibliographystyle{IEEEtranS}
\bibliography{refs}

\end{document}